\documentclass[11pt]{article}
\usepackage{amsmath,amssymb,color,graphics,epsfig,cite}

\usepackage{amsfonts}

\newcommand{\be}{\begin{equation}}
\newcommand{\ee}{\end{equation}}
\newcommand{\bea}{\setlength\arraycolsep{2pt} \begin{eqnarray}}
\newcommand{\eea}{\end{eqnarray}}
\newcommand{\nn}{\nonumber}

\def\ft#1#2{{\textstyle{\frac{\scriptstyle #1}{\scriptstyle #2} } }}
\def\fft#1#2{{\frac{#1}{#2}}}

\def\0{{\sst{(0)}}}
\def\1{{\sst{(1)}}}
\def\2{{\sst{(2)}}}
\def\3{{\sst{(3)}}}
\def\4{{\sst{(4)}}}
\def\5{{\sst{(5)}}}
\def\6{{\sst{(6)}}}
\def\7{{\sst{(7)}}}
\def\8{{\sst{(8)}}}
\def\sst#1{{\scriptscriptstyle #1}}

\begin{document}

\begin{center}
{\Large {\bf Horndeski Gravity as $D\rightarrow4$ Limit of Gauss-Bonnet}
}

\vspace{20pt}

{\large H. L\"{u} and Yi Pang}

\vspace{10pt}

{\it Center for Joint Quantum Studies and Department of Physics,\\
School of Science, Tianjin University, Tianjin 300350, China}

\vspace{40pt}

\underline{ABSTRACT}
\end{center}

We propose a procedure for the $D\rightarrow 4$ limit of Einstein-Gauss-Bonnet (EGB) gravity that leads to a well defined action principle in four dimensions. Our construction is based on compactifying $D$-dimensional EGB gravity on a $(D-4)$-dimensional maximally symmetric space followed by redefining the Gauss-Bonnet coupling $\alpha\rightarrow \frac{\alpha}{D-4}$. The resulting model is a special scalar-tensor theory that belongs to the family of Horndeski gravity. Static black hole solutions in the scalar-tensor theory are investigated. Interestingly, the metric profile is independent of the curvature of the internal space and coincides with the $D\rightarrow 4$ limit of the usual EGB black hole with the unusual Gauss-Bonnet coupling $\frac{\alpha}{D-4}$. The curvature information of the internal space is instead encoded in the profile of the extra scalar field. Our procedure can also be generalized to define further limits of the Gauss-Bonnet combination by compactifying the $D$-dimensional theory on a $(D-p)$-dimensional maximally symmetric space with $p\leq 3$. These lead to different $D\rightarrow 4$ limits of EGB gravity as well as its $D\rightarrow 2,3$ limits.

\vfill {\footnotesize  mrhonglu@gmail.com}


\thispagestyle{empty}
\pagebreak

\section{Introduction}

Nearly half a century ago, Lovelock showed that Einstein gravity could be extended by an infinite series of higher curvature terms such that the equations of motion remain second order \cite{L1}. The simplest such extension is Einstein-Gauss-Bonnet (EGB) gravity in dimensions
$D\ge 4$
\be
\label{EGB}
S_D=\frac{1}{16\pi G_D}\int d^Dx\,\sqrt{-g}\Big[R-2\Lambda_0+\alpha(R_{\mu\nu\rho\sigma}R^{\mu\nu\rho\sigma}-
4R_{\mu\nu}R^{\mu\nu}+R^2 )\Big]\,.
\ee
When $D=4$, the Gauss-Bonnet combination becomes a topological invariant and captures the global property of spacetime in the quantum gravity path integral. In $D>4$, the EGB gravity obeys Birkhoff's theorem and admits analytical static spherically symmetric solutions  \cite{EGBBH}
\begin{equation}
ds^2 = - f(r) dt^2 + f^{-1}(r) dr^2 + r^2 d \Omega_{D-2}^2\,,
\label{EGBSphSol}
\end{equation}
\begin{equation}
f_\pm(r) = 1 + \ft{r^2}{2 (D-3) (D-4) \alpha} \left[ 1 \pm
\sqrt{1 + \ft{64\pi G_{D} (D-3) (D-4) \alpha { M}}{(D-2) \Omega_{D-2} r^{D-1}}
+ \ft{8 (D-3) (D-4) \alpha \Lambda_0}{(D-1) (D-2)} }\right],
\label{eqn:MetricFunc}
\end{equation}
where $d\Omega_{D-2}^2$ is the metric of an $(D-2)$-dimensional unit sphere, and $\Omega_{D-2}$ is its volume. In view of the unitarity constraint $\alpha>0$ \cite{Cheung:2016wjt}, only solutions with $f_-$ correspond to black holes \cite{wheeler}. EGB gravity also admits topological black hole solutions with non-sphere horizon topology \cite{cai}. Thermodynamics of EGB black holes has been widely studied \cite{MS,cai,Cvetic:2001bk,Castro:2013pqa}. The black hole entropy computed using Iyer-Wald formula \cite{Wald:1993nt,Iyer:1994ys} exhibits a simple deviation from the Bekenstein-Hawking area term
\be
\label{entropy1}
S_{\rm BH}=\frac{\Omega_{D-2}r^{D-2}_+}{4G_D}\left(1+2\alpha(D-2)(D-3)r^{-2}_{+}\right)\,,
\ee
where $r_+$ is the radius of the outermost horizon. The recently revived interests on EGB gravity is based on
the observation that a limiting procedure defined as \cite{GL} \footnote{In fact, this limiting procedure is widely used in the context of conformal anomaly in the derivation of dilation effective action. See e.g. \cite{HHJ} for more discussions.}
\be
\label{limit}
\alpha\rightarrow \frac{\alpha}{D-4}\,,
\ee
followed by taking $D\rightarrow 4$ appears to retain
non-trivial effects from the Gauss-Bonnet term on the black hole solutions \cite{Konoplya:2020bxa}.
However, if the action \eqref{EGB} with $\alpha$ replaced by $\alpha/(D-4)$ is to be treated as the action of the limiting theory,
one immediately encounters a problem. From \eqref{entropy1}, it is evident that the entropy of the
black hole solutions obtained via the limiting procedure diverges as $D\rightarrow 4$. The standard thermodynamic relation between free energy and entropy also implies divergence in the on-shell Euclidean action. Thus the naive limiting procedure cannot properly describe topologically nontrivial solutions. In order to obtain a finite entropy for the limiting black hole solution, a diffeomorphism invariant regularisation is indispensable. Moreover, without a well defined local action, it is unclear how to count the dynamical degrees of freedom at fully nonlinear level using Hamiltonian analysis.

With the above questions in mind, we propose an action principle that defines the limit more concisely. Our procedure begins with the compactification of $D$-dimensional EGB gravity on a maximally symmetric space of $(D-4)$ dimensions, keeping only the breathing mode characterizing the size of the internal space. This Kaluza-Klein reduction ansatz is guaranteed be consistent in that the field equations of the four dimensional theory are compatible with those in higher dimensions. After removing a total derivative term, the limit \eqref{limit} can be smoothly applied, resulting a finite action taking the form of some special Horndeski gravity \cite{H} or generalized Galileons \cite{Deffayet:2009mn,VanAcoleyen:2011mj}, i.e.~a non-minimally coupled scalar-tensor theory with at most second order field equations. The limiting theory includes an extra scalar degree of freedom in addition to the spin-2 degrees of freedom. Static black holes in this particular Horndeski theory are investigated. Interestingly, the $D\rightarrow 4$ limit of the EGB black holes \eqref{eqn:MetricFunc} obtained through \eqref{limit} emerge as solutions independent of the curvature of the ``internal'' space on which original EGB gravity is compactified. The curvature of the ``internal'' space is instead recorded in the profile of the extra scalar field. The black hole entropy computed by applying Iyer-Wald formula to the new action is finite and correctly reproduces the finite part of the entropy of the EGB black holes in the limit \eqref{limit}. Our action can therefore account for the Euclidean path-integral for the topologically nontrivial solutions.

\section{Limits of Einstein-Gauss-Bonnet gravity in $D<5$}

We begin with the parametrization of $D$-dimensional metric
\be
ds^2_D=ds^2_p+e^{2\phi}d\Sigma^2_{D-p,\,\lambda}\,,
\ee
where the breathing scalar $\phi$ depends only on the external $p$-dimensional coordinates, $ds_p^2$ is the $p$-dimensional line element and $d\Sigma^2_{D-p,\,\lambda}$ is the line element of an internal maximally symmetric space of $(D-p)$ dimensions with curvature tensor
\be
R_{abcd}= \lambda(g_{ac}g_{bd}-g_{ad}g_{bc})\,.
\ee
The reduction ansatz above is the simplest nontrivial assumption we can make in order to focus on the S-wave excitation of the metric. The action (\ref{EGB}) then consistently reduces to the $p$-dimensional action\footnote{Here we corrected a few typos in the results presented in \cite{CGK,C}.}
\bea
\label{p-action}
&&S_p=\frac{1}{16\pi G_p}\int d^px\sqrt{-g}e^{(D-p)\phi}\Bigg\{R
-2\Lambda_0+(D-p)(D-p-1)\big((\partial\phi)^2+\lambda e^{-2\phi}\big)\cr
&&
+\alpha\Big({\rm GB}-2(D-p)(D-p-1)\left[2G^{\mu\nu}\partial_\mu\phi\partial_\nu\phi-\lambda Re^{-2\phi}\right]\cr
&& -(D-p)(D-p-1)(D-p-2)\left[2(\partial\phi)^2\Box\phi+(D-p-1)((\partial\phi)^2)^2\right]\cr
&&+(D-p)(D-p-1)(D-p-2)(D-p-3)\left[2\lambda (\partial\phi)^2e^{-2\phi}+\lambda^2e^{-4\phi}\right]\Big)\Bigg\}\,,
\eea
where $G_{\mu\nu}$ is the Einstein tensor. In this paper, we restrict to the case $p\leq 4$ in which the Gauss-Bonnet combination is either topological or
identically 0. Therefore we can add
\be\label{GBterm}
-\frac{\alpha}{16\pi G_p}\int d^px\sqrt{-g}\, {\rm GB}
\ee
to the action \eqref{p-action}, leaving the field equations inert. We now redefine the Gauss-Bonnet coupling by $\alpha\rightarrow \frac{\alpha}{D-p}$  and take the limit $D\rightarrow p$, yielding
\bea
\label{limitS}
S_p&=&\int d^px\sqrt{-g}\Big[R
-2\Lambda_0+\alpha\Big(\phi\, {\rm GB}+4G^{\mu\nu}\partial_\mu\phi\partial_\nu\phi
-2\lambda Re^{-2\phi}-4(\partial\phi)^2\Box\phi\nn\\
&&+2((\partial\phi)^2)^2-12\lambda(\partial\phi)^2e^{-2\phi}-6\lambda^2e^{-4\phi}\Big)\Big]\,.
\eea
The action above defines a limit for EGB not only in $D=4$ but also in $D<4$. Similar to the Gauss-Bonnet combination, the coefficients in the limiting action \eqref{limitS} is independent of the spacetime dimensions. (The theory can thus be promoted to general higher dimensions, at the price that the connection to the EGB theory is lost.) When $\lambda=0$, corresponding to flat ``internal'' space, the theory is invariant under a constant shift of $\phi$ and is equivalent to 4-dimensional dilation effective action appearing in the study of conformal anomlay \cite{HHJ}.

In fact, the action \eqref{p-action} admits two more classes of lower-$D$ limits. The first one is to
scale $\alpha\rightarrow \frac{\alpha}{D-p-1}$ followed by $D\rightarrow p+1$.  This leads to theories in $p\le 3$ dimensions:
\bea
S_p&=&\int d^px\sqrt{-g}e^{\phi}\Big[R
-2\Lambda_0\nn\\
&&+\alpha\Big(-4G^{\mu\nu}\partial_\mu\phi\partial_\nu\phi
+2\lambda Re^{-2\phi}+2(\partial\phi)^2\Box\phi+4\lambda(\partial\phi)^2e^{-2\phi}+2\lambda^2e^{-4\phi}\Big)\Big]\,.
\eea
The second one exists only in $p\leq 2$ and is defined by
\bea
\label{limitS3}
S_p&\rightarrow& S_p
-\frac{\alpha\lambda (D-p)(D-p-1)}{8\pi G_p}\int d^px\sqrt{-g}R\,,\quad \alpha\rightarrow \frac{\alpha}{D-p-2}\,,\quad D\rightarrow p+2\,,
\eea
which results in
\bea
S_p&=&\int d^px\sqrt{-g}e^{2\phi}\Big[R
-2\Lambda_0+2(\partial\phi)^2+2\lambda e^{-2\phi}+2\alpha\Big(2\lambda \phi Re^{-2\phi}-2(\partial\phi)^2\Box\phi\nn\\
&&-((\partial\phi)^2)^2-2\lambda(\partial\phi)^2e^{-2\phi}-\lambda^2e^{-4\phi}\Big)\Big]\,.
\eea
Both classes include new $D\rightarrow 4$ limits that actually yield lower $p$-dimensional theories.

\section{Black hole solutions}

We now study black hole solutions in four dimensions. We adopt the ansatz for the metric and scalar field $\phi$
\be
\label{ansatz}
ds_4^2 = - e^{-2\chi(r)} f(r) dt^2 + \fft{dr^2}{f(r)} + r^2 d\Omega_{2,k}^2\,,\quad d\Omega_{2,k}^2 = \fft{dx^2}{1-k x^2} + x^2 d\varphi^2\,,\quad \phi=\phi(r)\,,
\ee
where $k=-1,0,1$. Upon substituting the ansatz to $S_4$ in (\ref{limitS}), we obtain the effective Lagrangian for variables $(f,\chi,\phi)$
\bea
L_{\rm eff} &=&e^{-\chi}\Big[2(k-\Lambda_0 r^2 - f- r f') +\ft23\alpha\phi'\Big(
3 r^2 f^2 \phi '^3+2 r f \phi '^2 \left(-r f'+2 r f \chi '-4 f\right)\nn\\
&&-6 f \phi ' \left(-r f'+2 r f \chi '-f+k\right)-6 (f-k) \left(f'-2 f \chi '\right)\Big)\nn\\
&& + 4\alpha \lambda e^{-2\phi} \Big(r^2 f' \phi '-2 r^2 f \chi ' \phi '-3 r^2 f \phi '^2+r f'+f-k\Big)-6\alpha\lambda^2 r^2 e^{-4\phi}\Big]\,,
\eea
where some total derivative terms were dropped. The equations of motion can be simplified and $(f,\phi)$ form a closed subsystem of equations
\be
f' = f'(f,\phi', \lambda e^{-2\phi})\,,\qquad \phi''=\phi''(f,\phi', \lambda e^{-2\phi})\,,\label{eom}
\ee
and $\chi$ is given by
\be
\chi' = \fft{1}{f} \Big((r\phi'-1)^2 f - \lambda r^2 e^{-2\phi} -k\Big) P(f,\phi', \lambda e^{-2\phi})\,,
\label{chip}
\ee
where $P$ is a complicated expression of rational polynomials of functions $(f,\phi')$ and $\lambda e^{-2\phi}$. When $\lambda=0$, the equations involve $(\phi',\phi'')$ only.

\subsection{$\chi=0$}

It follows from (\ref{chip}) that $\chi=0$ is a consistent truncation. In this case, $f$ and $\phi$ satisfy
\bea
&&r^2 \left(r f'+f+\Lambda _0 r^2-k\right)+\alpha  (f-k) \left(f-k -2 r f'\right)=0\,,\\
&&(r \phi'-1)^2 f = \lambda r^2 e^{-2\phi} +k\,.\label{scalareom}
\eea
The most general solution for $f$ is
\be
f_\pm =k + \fft{r^2}{2\alpha} \Big(1 \pm \sqrt{1 + \ft43\alpha \Lambda_0 + \fft{8\alpha M}{r^3}}\Big)\,.
\ee
This is precisely the $D\rightarrow 4$ limit of the EGB black hole discussed in the introduction which also arises in the conformal anomaly inspired gravity \cite{Cai:2009ua}. This is expected in view of the consistency of the dimensional reduction which implies \cite{Ma:2020ufk}
\be
R-4\Lambda_0+\ft12\alpha {\rm GB}=0
\ee
should hole as a consequence of the trace part of the EGB field equation. This equation together with the $\chi=0$ ansatz determines the metric to take the expression given in (20). Different from the higher dimensional black holes \eqref{eqn:MetricFunc} in which the singularity is spacelike, the 4-dimensional metric admits two horizons and thus the singularity is time-like. Note that the metric is independent of the parameter $\lambda$, but the scalar $\phi$ is. For generic $\lambda$, the solution for $\phi$ is given by
\be
\label{scalarsol}
\phi_\pm = \log \ft{r}{L} +\log \Big(\cosh (\sqrt{k}\,\psi) \pm \sqrt{1+\lambda\,L^2k^{-1}} \sinh(\sqrt{k}\,\psi)\Big)\,,\quad \psi = \int_{r_+}^r \fft{du}{u\sqrt{ f(u)}}\,,
\ee
where $L$ is an arbitrary integration constant. We note that since the left-hand side of \eqref{scalareom} is non-negative, $\lambda$ and $k$ cannot be  simultaneously negative. When $\lambda\,,k>0$, $\phi_+$ diverges at an intermediate radius value between the horizon and infinity and thus should be discarded. When $k<0$, the expression \eqref{scalarsol} can still be real once $\sqrt{1+\lambda\,L^2k^{-1}}$ is replaced by ${\rm i} \sqrt{\lambda\,L^2|k^{-1}|-1}$. Although in general $\phi_\pm$ has $\log(r)$ divergence asymptotically, the solutions are still well defined since only $\phi'$ and $e^{-2\phi}$, which are convergent, appear in the theory, giving rise to no modification to the black hole mass.  For $\lambda=0$ and $k=1$, $\phi_-$ vanishes asymptotically, which we shall revisit presently.

Using the Iyer-Wald formula, the entropy of the black hole ($k=1$) can be obtained as
\be
\label{entropy2}
S=\ft{1}{G_4} \left(\pi r_+^2 + 4\alpha\pi \log \ft{r_+}{L}\right)\,,
\ee
where we have absorbed an inessential constant contributed from $\alpha \lambda Re^{-2\phi}$ term to $L$. (The $G_{\mu\nu} \partial^\mu \phi \partial^\nu \phi$ does not contribute to any entropy \cite{Feng:2015oea}.) It is easy to see that the first law of black hole thermodynamics $dM=TdS$ is satisfied if $L$ is held fixed. Since $L$ is an arbitrary constant, there seems to be an ambiguity in the black hole entropy. In fact, similar ambiguity is also present in the $D\rightarrow 4$ limit of entropy \eqref{entropy1}. To see this, we first parameterize the $D$-dimensional Netwon's constant in terms of the $4$-dimensional constant as $G_D=G_4 \ell^{D-4}$ where another length scale $\ell$ is introduced to balance the dimension difference between $G_D$ and $G_4$. Now as $D\rightarrow 4$, The entropy \eqref{entropy1} becomes
\be
S_{\rm BH}\big|_{D\rightarrow 4}=\ft{c}{D-4}+\ft{1}{G_4} \left(\pi r_+^2 + 4\pi\alpha \log \ft{r_+}{\ell}\right)\,,
\ee
where the reference scale $\ell$ is not determined from first principle. By comparing the expression above with the entropy computed using Iyer-Wald formula \eqref{entropy2}, it seems that within the scalar-tensor theory, the reference scale $\ell$ can be understood as the integration constant arising from solving the scalar field equation.

When the mass parameter $M$ vanishes, the metric becomes maximally symmetric spacetimes with
\be
f=-\ft13 \Lambda_{\rm eff} r^2 + k\,,\qquad \Lambda_0 = \Lambda_{\rm eff} + \ft13\alpha \Lambda_{\rm eff}^2\,.
\ee
However, the full spacetime symmetry can be broken since $\phi$ can be $r$-dependent.  The true spacetime vacuum is when $\phi=\phi_0$ is constant, satisfying
\be
\lambda=-\ft13\Lambda_{\rm eff}\, e^{2\phi_0}\,.
\ee
Thus Minkowski or (A)dS vacua require vanishing and non-vanishing $\lambda$ respectively. For $\Lambda_{\rm eff}=0$ and hence $k=1$, the $\phi_-$ solution is asymptotical to Minkowski spacetime, whilst the $\phi_+\sim \log(r)$  is not, even though the black hole metrics are identical. Note that when $\lambda=0=k$, we have AdS spacetime in planar coordinates with $\phi=\log r$, giving rise to a vacuum with Poincar\'e and scaling invariance, leading to a possible dual of scaling invariant but not conformal invariant quantum field theory \cite{Li:2018rgn}.

\subsection{$\chi\neq0$}

The equations become much more complicated and it is unlikely to have analytical solutions with $\chi\ne 0$.
We shall study this issue using numerical approach.  For simplicity, we consider the $\Lambda_0=0=\lambda$ case, where the $\chi=0$ black hole was constructed earlier.  Asymptotically, the leading falloffs read
\be
f=1-\frac{2 M}{r}+\frac{4 \alpha  M^2}{r^4}+\cdots\,,
\ee
with the two $\phi$ branches
\be
\phi_+'=\frac{2}{r}+\frac{M}{r^2}+\frac{3M^2}{2 r^3}+\frac{5M^3}{2 r^4}+\cdots\,,\qquad
\phi_-'=-\frac{M}{r^2}-\frac{3M^2}{2 r^3}-\frac{5M^3}{2 r^4}+\cdots\,.
\ee
It is clear that only the second branch yields an asymptotically-flat black hole. (See earlier discussions.) For non-vanishing $\chi$, the asymptotic behaviors are
\be
f=1-\fft{2M}{r} - \fft{12\alpha M^2}{r^4}+ \cdots\,,\quad
\phi'=\fft{M}{r^2} + \fft{2M^2}{r^3} + \fft{4M^3}{r^4}+ \cdots\,,\quad
\chi' = \fft{16\alpha M^2}{r^5} + \cdots\,.\label{newasym}
\ee
The leading falloffs of $\phi_-'$ and $\phi'$ are the same but in opposite signs, and the $\chi\ne0$ solution is therefore asymptotically flat. It should be pointed out that with our choice of positive $\alpha$, the metric with $\chi'>0$ would violate the null energy condition in the framework of Einstein gravity with minimally coupled matter, and hence the above metric cannot be constructed in Einstein gravity with a normal matter energy-momentum tensor.

\begin{figure}[ht]
  \centering
  \includegraphics[width=0.4\textwidth]{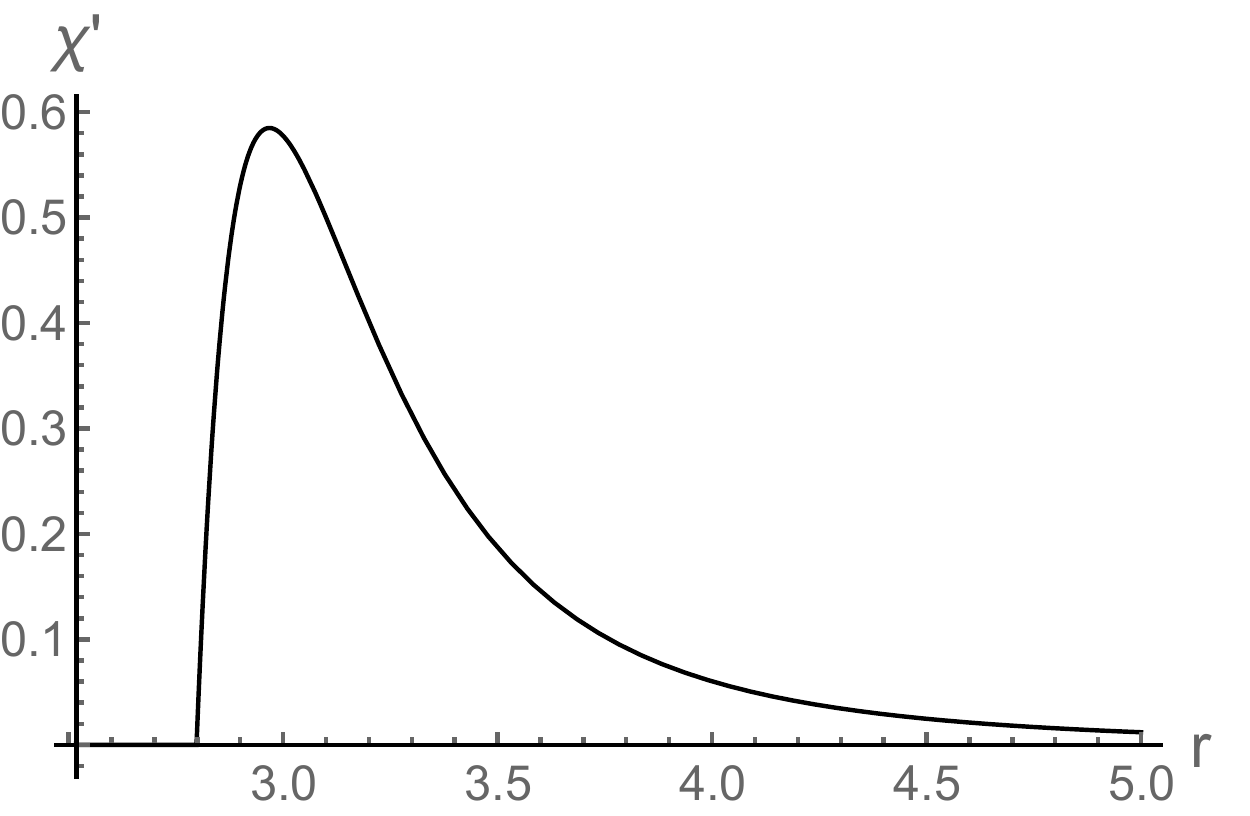}\ \ \includegraphics[width=0.5\textwidth]{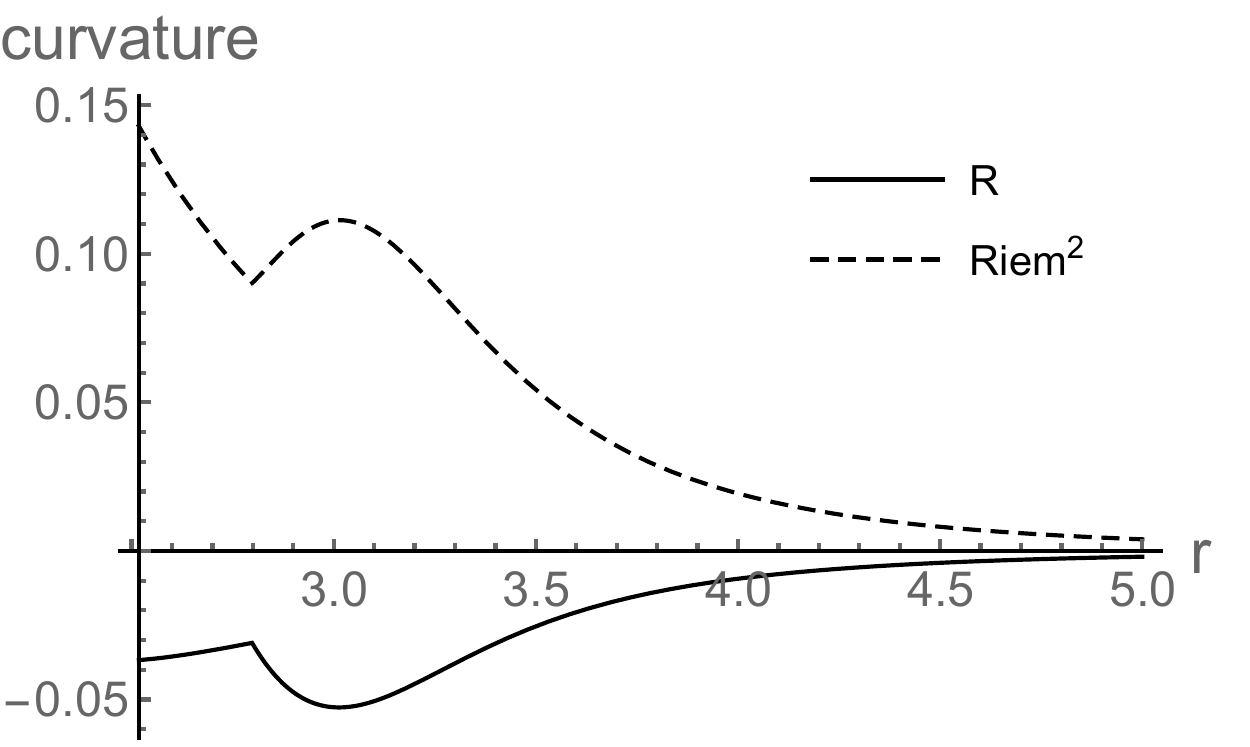}
  \caption{\small The left shows the the function $\chi'(r)$, which vanishes from the horizon $r_+=2.5181$ to $r_1=2.798$.  The maximum is at $r_2=2.968$.  The right gives the Ricci and Riemann$^2$ curvatures that develops cusps at $r_1$. The cusps imply $\delta$-function singularities that appear in the covariant derivatives of the curvature tensors.   The solution has $\alpha=1$ and $M=1$. } \label{chifig}
\end{figure}

\begin{figure}[ht]
  \centering
  \includegraphics[width=0.4\textwidth]{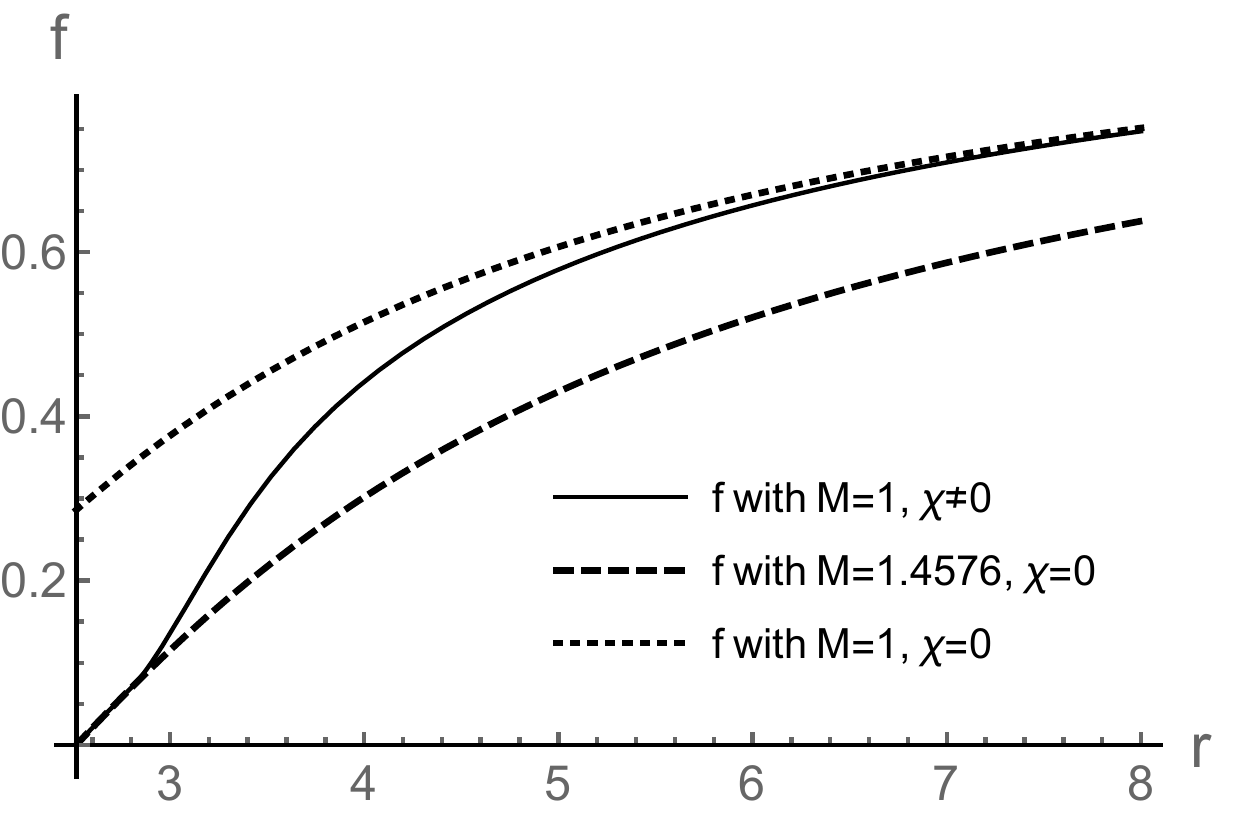}\ \ \  \includegraphics[width=0.4\textwidth]{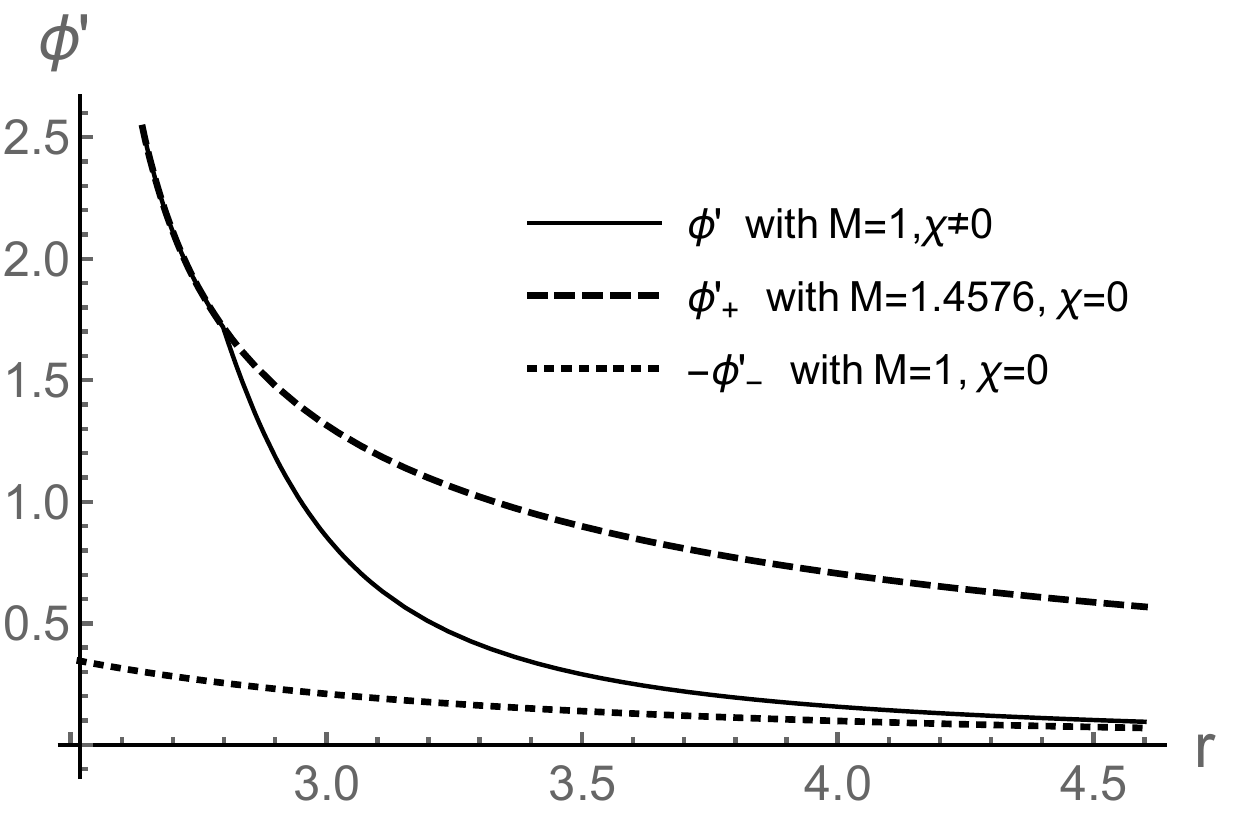}
  \caption{\small The left is the $f$ solution for $\chi\ne0$, with $\alpha=1$ and $M=1$.  It joins
  to the $\chi=0$ solution of mass $M=1.4576$ whose horizon is also at $r_+=2.5181$. Asymptotically,
   it approaches the known $\chi=0$ black hole with $M=1$, by construction.
   The right plot is the corresponding $\phi'$.  It joins the $\phi_+'$ with $M=1.4576$ near the horizon and approaches $-\phi_-'$ at infinity, which falls off as $1/r^2$.} \label{fphipfig}
\end{figure}

One can naturally ask whether there exists some horizon structure that integrates out to the asymptotic region (\ref{newasym}).  After a careful analysis, we find that regularity requires that $\chi=0$ in the vicinity of the horizon.  This seems to suggest that the asymptotic structure (\ref{newasym}) is not related to a black hole.  Numerical results however reveal something more intriguing. It follows from (\ref{eom}) that $(f,\phi)$ form a closed system of equations and can be solved within themselves, and we can then read off  $\chi'$ from (\ref{chip}).  For a concrete example, we take $\alpha=1$ and $M=1$ for numerical analysis, and we present $\chi'$ in the left plot of Fig.~\ref{chifig}.
We integrate from large $r$ to 0, and we see that as $r$ runs from the asymptotic infinity to the middle, $\chi'$ increases and reaches the maximum at $r_2=2.968$ and then decreases to zero at $r_1=2.798$. What is intriguing is that once $\chi'$ vanishes, it stays zero and the solution begins to run as the $\chi=0$ solution of the $\phi_+$ branch with $M=1.4576$, until it reaches the horizon at $r_+=2.5181$, as indicated by the plots of $f$ and $\phi'$ in Fig.~\ref{fphipfig}.
The joining is smooth for the $f$ function, but there is a cusp for the $\phi'$ function. Asymptotically, $\phi'$ for the $\chi\ne0$ solution matches the $-\phi_-'$ of the mass $M=1$, with the falloff $M/r^2$, but it slumps into $\phi_+'$ of the mass $M=1.4576$ in the vicinity of the horizon.  The curvatures develop cusps at $r_1$, as shown in the right plot of Fig.~\ref{chifig}. The cusps imply that there are $\delta$-function singularities in covariant derivatives of the curvature, e.g.~$\Box R$.  These singularities are naked, but much milder than the power-law singularity at $r=0$. This new $\chi\ne0$ solution captures the essence of the nonlinear dynamics of the scalar field in Horndeski gravity as the $D\rightarrow 4$ limit of the EGB gravity.

\section{Conclusions}

In this paper, we proposed a mathematically more rigorous definition for the $D\rightarrow 4$ limit of EGB gravity. It involves compactifying $D$-dimensional EGB gravity on a $(D-4)$-dimensional maximally symmetric space, subtracting a (divergent) total derivative term and redefining the Gauss-Bonnet coupling $\alpha\rightarrow \frac{\alpha}{D-4}$. The resulting model is a special scalar-tensor theory that belongs to the family of Horndeski gravity. We also studied static black hole solutions in this theory and observed that the metric profile coincides with that of EGB black holes under the limit \eqref{limit}, unaffected by the curvature of the internal space on which the original EGB is compactified. The profile of the scalar field does depend on the internal curvature. We also obtained new black holes whose EGB origin remains to be investigated.

Our procedure can be generalized to define further limits of Gauss-Bonnet combination by compactifying the $D$-dimensional theory on a $(D-p)$-dimensional maximally symmetric space with $p\leq 3$ together with an appropriate redefinition of its coupling. These lead to different $D\rightarrow 4$ limits of EGB gravity as well as its $D\rightarrow 2,3$ limits which are described by lower-dimensional Horndeksi models, of which the black hole solutions are not well-understood and deserve future research. We believe that applying our procedure to general Lovelock theories, one can also obtain non-trivial lower dimensional limits described by more general Horndeski models in various dimensions. Finally, it should be interesting to investigate the cosmological implication of the particular four-dimensional Horndeski model obtained in the paper.

\section*{Acknowledgement}
Y.P. has benefited from many informative discussions with Stanley Deser who also encouraged the authors to publish this paper. We are also grateful to Bayram Tekin for useful discussions. H.L.~is supported in part by NSFC (National Natural Science Foundation of China) Grants No.~11875200 and No.~11935009.

\end{document}